# SCALING BEHAVIOUR AND COMPLEXITY OF THE PORTEVIN-LE CHATELIER EFFECT

A. Sarkar and P. Barat

Variable Energy Cyclotron Centre

1/AF Bidhan Nagar, Kolkata 700064, India

**Abstract:**

The plastic deformation of dilute alloys is often accompanied by plastic instabilities due to dynamic strain aging and dislocation interaction. The repeated breakaway of dislocations from and their recapture by solute atoms leads to stress serrations and localized strain in the strain controlled tensile tests, known as the Portevin-Le Chatelier (PLC) effect. In this present work, we analyse the stress time series data of the observed PLC effect in the constant strain rate tensile tests on Al-2.5%Mg alloy for a wide range of strain rates at room temperature. The scaling behaviour of the PLC effect was studied using two complementary scaling analysis methods: the finite variance scaling method and the diffusion entropy analysis. From these analyses we could establish that in the entire span of strain rates, PLC effect showed Levy walk property. Moreover, the multiscale entropy analysis is carried out on the stress time series data observed during the PLC effect to quantify the complexity of the distinct spatiotemporal dynamical regimes. It is shown that for the static type C band, the entropy is very low for all the scales compared to the hopping type B and the propagating type A bands. The results are interpreted considering the time and length scales relevant to the effect.

**Keywords:** Dynamic strain aging, Standard deviation analysis, Diffusion entropy analysis, Multiscale entropy analysis

## 1. Introduction

The theoretical value of the shear strength of metals is much higher than the observed shear strength - this discrepancy remained unexplained until 1934 when Taylor, Orowan and Polanyi [1-3] introduced the concept of dislocation. The dislocation is the most important two-dimensional line defect in crystals which is responsible for the phenomenon of slip, by which most metals deform plastically. The existence of the dislocation was confirmed experimentally in 1950's [4]. At the microscopic scale the strength of a crystal derives from the motion and interaction of the dislocations. The ability of a crystalline material to deform plastically largely depends on the ability for the dislocations to move within the material. Therefore, impeding the movement of dislocations results in the strengthening of the material. There are number of ways to hinder dislocation movement, which include: (a) controlling the grain size: reducing continuity of atomic planes (b) strain hardening: creating and tangling dislocations (c) solid solution strengthening: introducing point defect to pin dislocation. In this paper we will be concerned with the third mechanism. The introduction of solute atoms into the solid solution in the solvent-atom lattice invariably produces an alloy which is stronger than the parent one. There are two types of solid solutions depending upon the position the solute atoms occupy in the host matrix: interstitial and substitutional. The added solute atoms are point defects to the lattice and they have their own strain fields. These strain fields interact with the strain field of the dislocations through various mechanisms like elastic interaction, modulus interaction, electronic interaction etc. and help to restrict the movement of the dislocations.

In principle the motion of a single dislocation in a perfect crystal is a simple phenomenon. However, in a real crystal many cooperative dislocation effects may give rise to unusual and complex plastic properties which are in general designated as plastic instabilities. The Portevin-Le Chatelier (PLC) effect is one of the prominent examples of the plastic instabilities.

Many dilute solid solutions of technological importance, both interstitial and substitutional, exhibit the PLC effect. This phenomenon is observed in a certain range of temperature and strain rate and consists in the deformation being inhomogeneous, dominated by successive strain localization events. At a certain value of the plastic strain, the critical strain $\varepsilon_c$, the character of the deformation changes from homogeneous to inhomogeneous, which is indicated by the occurrence of serrations on the stress strain curve which ultimately leads to band-type macroscopic surface markings. These limit the potential application of material when good surface quality is required. The physical origin of the PLC effect is attributed to dynamic strain aging (DSA) [5-10], a generic term representing a host of small-scale phenomena associated with the interaction of dislocation and solute atoms. DSA leads to the negative strain rate sensitivity (SRS) [9-12]. In fact, the PLC effect is a direct manifestation of the negative SRS, which gives rise to material instability. The ratio of the flow stress variation corresponding to an imposed strain rate variation is the measure of the SRS. If this becomes negative, conditions exist for the macroscopic observation of the PLC effect.

Savart and Mason [13,14] discovered this discontinuous behavior of the plastic deformation in the 19$^{th}$ century. A first systematic and detailed study was carried out by Le Chatelier [15] in 1909 on mild steel samples deformed at high temperatures and later by Portevin and Le Chatelier on Duralumin in 1923 [16]. The preliminary explanation to the PLC effect was suggested by Le Chatelier prior to the advent of dislocation theory. The introduction of the concept of dislocation revolutionized the study of plastic flow hence that of the PLC effect.

In the last few decades several studies have been carried out on the PLC effect and several models have also been proposed. The first model was due to Cottrell [17] which is based on the pinning and unpinning of dislocations by the solute atoms. In this model the basic assumption is that the solute atoms have sufficient mobility to follow the dislocations as they move continuously through the lattice. This process leads to an increased friction force for mobile dislocations. It was later realized [9] that solute atoms do not have sufficient mobility to follow moving dislocations and aging should take place while dislocations are arrested at obstacles such as at forest dislocations. Van den Beukel [7] developed this idea into a quantitative model of DSA. The basic mechanism envisioned is clustering of the solute atoms at the mobile dislocations temporarily arrested at the obstacles through the bulk diffusion or by pipe diffusion. Dislocations overcome the barrier by the aid of thermal activation and jump at the next obstacles at high velocity.

Apart from these theoretical studies the PLC effect has also been a topic of several experimental investigations in the last few decades [18-25]. The PLC effect is most conveniently studied through tensile tests either at constant applied stress rate ('soft' device) or at constant strain rate ('hard' device): within certain ranges of the control variable (stress and strain rate) and deformation temperature. In uniaxial loading with a constant imposed strain rate, the effect manifests itself as serrations in the stress-time (or strain) curves. This is associated with the repeated generation and propagation of plastic deformation bands. The bands mark the region of appreciable plastic deformation. Each stress drop is associated with the nucleation of a band of localized plastic deformation.

In polycrystals three types of the PLC effect are traditionally distinguished on the qualitative basis of the spatial arrangement of localized deformation loads and the particular appearance of deformation curves [26-30]. Three generic types of serrations: type A, B and C occur depending on the imposed cross-head velocity i.e. strain rate. For sufficiently large strain rate, type A serrations are observed. In this case, the bands are continuously propagating and highly correlated. The associated stress drops are small in amplitude [31-36].

If the strain rate is lowered, type B serrations with relatively larger amplitude occur around the uniform stress strain curve. These serrations correspond to intermittent band propagation. The deformation bands are formed ahead of the previous one in a spatially correlated manner and give rise to regular surface markings [31-36].

For even smaller strain rate, bands become static. This type C band nucleates randomly in the sample leading to large saw-tooth shaped serration in the stress strain curve and random surface markings [31-36].

The PLC effect has been extensively studied over the last several decades with the goal being to achieve a better understanding of the small-scale processes and of the multiscale mechanisms that link the mesoscale DSA to the macroscale PLC effect. The technological goal is to increase the SRS to positive values in the range of temperatures and strain rates relevant for industrial processes. This would ensure material stability during processing and would eliminate the occurrence of the PLC effect.

Beyond its importance in metallurgy, the PLC effect is a paradigm for a general class of nonlinear complex systems with intermittent bursts. The succession of plastic instabilities shares both physical and statistical properties with many other systems exhibiting loading-unloading cycles e.g. earthquake. PLC effect is regulated by interacting mechanisms that operate across multiple spatial and temporal scales. The output variable (stress) of the effect exhibits complex fluctuations which contains information about the underlying dynamics.

Due to a continuous effort of numerous researchers, there is now a reasonable understanding of the mechanisms and manifestations of the PLC effect. A review of this field can be found in Ref. [33]. The possibility of chaos in the stress drops of PLC effect was first predicted by G. Ananthakrishna *et.al* [37] and latter by V. Jeanclaude *et. al*. [38]. This prediction generated a new enthusiasm in this field. In last few years, many statistical and dynamical studies have been carried out on the PLC effect [39-45]. Analysis revealed two types of dynamical regimes in the PLC effect. At medium strain rate (type B) chaotic regime has been demonstrated [39-41], which is associated with the bell-shaped distribution of the stress drops. For high strain rate (type A) the dynamics is identified as self organized criticality (SOC) with the stress drops following a power law distribution [42-44]. The crossover between these two mechanisms has also been a topic of intense research for the past few years [42-45]. It is shown that the crossover from the chaotic to SOC dynamics is clearly signaled by a burst in multifractality [45]. M. Lebyodkin *et. al.* [40, 41] have studied the spatio-temporal dynamics of the PLC effect in detail. In one of their works they have also proposed the PLC effect to be a candidate for modeling the earthquake statistics. Other researchers like M.S.Bharathi *et. al.* [43] and S. Kok *et. al.* [46] have studied the dynamical and chaotic behavior of the PLC effect.

Despite many dynamical studies, PLC effect remains an active area of research, with many important questions still open. This paper focuses on an alternate approach to establish the connection between the dislocation interactions and the stress fluctuations of the PLC effect. A link between these two phenomena is detected through a detailed scaling analysis of the time series data of the stress fluctuations during PLC effect in the different ranges of plastic instabilities. We also quantify the complexity of the PLC effect observed at different strain rate regime.

The physics of the PLC effect is described by the coupling of nonlinear equations. DSA is made up of the local variations in stress, strain and strain rate. Based on the early results, we anticipate that the dynamical equations for DSA are stochastic and nonlinear. However, we do not undertake the daunting task of deriving these equations here, but rather examine the time series data sets of stress and from the analyses, deduce some characteristics of the stress variations.

The general consensus that the dynamic strain aging is the cause behind the PLC effect suggested a discrete connection between the stress fluctuation and the band dynamics. We do not have system of primitive equation to describe the dynamics of the band, so we must extract as much information as possible from the data. We use the stress data recorded during the plastic deformation for our analysis. However, we do not analyze these data blindly but are guided by the nonlinear dynamics that would be produced by the observed intermittency in a time of band dynamics.

## 2. Experimental

5xxx class substitutional Aluminum alloys with Mg as the primary alloying element are model systems for the PLC effect studies. These alloys have wide technological applications due to their advantageous strength to weight ratio. They show good ductility and can be rolled to large reductions and processed in thin sheets and are being extensively used in beverage packaging and other applications. However, the discontinuous deformation behavior of these alloys at room temperature rule them out from many important applications like in automobile industry. These alloys exhibit the PLC effect for wide range of strain rates and temperatures. Under these deformation conditions the deformation of these materials localize in narrow bands which leave undesirable band-type macroscopic surface markings on the final products. These alloys are of substitutional type and Mg is substitutional in Al. The atomic radius of Mg is larger than that of Al and the radius mismatch between them is ~ 12%. As Mg atom is larger in size it acts as a center of dilatation when it substitutes Al atom in the Al matrix. We have carried out our studies on the PLC effect of the Al-2.5%Mg alloy at room temperature.

Tensile testing was conducted on flat specimens prepared from polycrystalline Al-2.5%Mg alloy. Specimens with gauge length, width and thickness of 25, 5 and 2.3 mm, respectively were tested in an INSTRON (model 4482) machine. All the tests were carried out at room temperature (300K) and consequently there was only one control parameter, the applied strain rate. To monitor closely its influence on the dynamics of the PLC effect, strain rate was varied from $7.56 \times 10^{-6}$ Sec$^{-1}$ to $1.92 \times 10^{-3}$ Sec$^{-1}$. The PLC effect was observed through out the range. The stress-time response was recorded electronically at periodic time intervals of 0.05 seconds. Fig. 1 shows the observed PLC effect in a typical stress-strain curve for strain rate $3.85 \times 10^{-4}$ Sec$^{-1}$. The inset shows the magnified view of stress-time variation of a typical region in it. In the varied strain rate we could observe type A, B and C bands as reported [28, 47].

## 3. Scaling Behaviour of the PLC effect

Scaling as a manifestation of the underlying dynamics is familiar throughout physics. It has been instrumental in helping scientists gain deeper insights into problems ranging across the entire spectrum of science and technology. Scaling laws typically reflect underlying generic features and physical principles that are independent of detailed dynamics or characteristics of particular models. Scale invariance has been found to hold empirically for a number of complex systems, and the correct evaluation of the scaling exponents is of fundamental importance in assessing if universality classes exist [48]. Scale invariance seems to be widespread in natural systems. Numerous examples of scale invariance properties can be found in the literature like earthquakes, clouds, networks etc [49-52].

The basic idea behind the study of the scaling behaviour of the PLC effect is to see whether the dislocation intermittency due to the DSA induces a scaling behaviour in the stress fluctuations. Such dynamical stochastic processes can be described by the generalizations of random walks. A Levy flight, for example, is such a process with diverging second moment [53]. A Levy walk, on the other hand, visits the same spatial sites as does a Levy flight, but each step takes a finite time and the second moment is finite. The time necessary to complete a step in a Levy walk is specified by an inverse power law waiting time distribution [53].

To study the scaling behavior of the PLC effect we make use of two complementary scaling analysis methods: the finite variance scaling method (FVSM) and the diffusion entropy analysis (DEA) [54-58]. The need for using these two methods to analyze the scaling properties of a time series is to discriminate the stochastic nature of the data: Gaussian or Levy [56]. Recently, Scafetta *et al.* [59] had shown that to distinguish between fractal Gaussian intermittent noise and Levy-walk intermittent noise, the scaling results obtained using DEA should be compared with that obtained from FVSM.

### 3. 1. FVSM and DEA

These methods are based on the prescription that numbers in a time series $\{\xi_i\}$ are the fluctuations of a diffusion trajectory; see Refs. [55, 60, 61] for details. Therefore, we shift our attention from the time series $\{\xi_i\}$ to probability density function (pdf) $p(x,t)$ of the corresponding diffusion process. Here x denotes the variable collecting the fluctuations and is referred to as the diffusion variable. The scaling property of $p(x,t)$ takes the form

$$p(x,t) = \frac{1}{t^\delta} F\left(\frac{x}{t^\delta}\right) \qquad (1)$$

In the FVSM one examines the scaling properties of the second moment of the diffusion process generated by a time series. One version of FVSM is the standard deviation analysis (SDA) [55], which is based on the evaluation of the standard deviation $D(t)$ of the variable $x$, and yields [55].

$$D(t) = \sqrt{\langle x^2;t \rangle - \langle x;t \rangle^2} \propto t^H \qquad (2)$$

The exponent $H$ is interpreted as the scaling exponent and is usually called the Hurst exponent. It is evaluated from the gradient of the fitted straight line in the log-log plot of $D(t)$ against $t$.

The DEA was developed [54] as an efficient way to detect the scaling and memory in time series for variables in complex systems. This procedure has been successfully applied to sociological, astrophysical and biological time series [54-57]. DEA focuses on the scaling exponent δ evaluated through the Shannon entropy $S(t)$ of the diffusion generated by the fluctuations $\{\xi_i\}$ of the time series [54, 55]. Here, the pdf of the diffusion process, $p(x,t)$, is evaluated by means of the sub trajectories $x_n(t) = \sum_{i=0}^{t} \xi_{i+n}$ with $n=0,1,..$ If the scaling condition of Eq. (1) holds true, it is easy to prove that the entropy

$$S(t) = -\int_{-\infty}^{\infty} p(x,t)\ln[p(x,t)]dx \qquad (3)$$

increases in time as

$$S(t) = A + \delta \ln(t) \qquad (4)$$

with

$$A = -\int_{-\infty}^{\infty} dy F(y)\ln[F(y)] = \text{Constant}, \qquad (5)$$

where $y = \frac{x}{t^\delta}$. Eq. (4) indicates that in the case of a diffusion process with a scaling pdf, its entropy $S(t)$ increases linearly with $\ln(t)$. The scaling exponent $\delta$ is evaluated from the gradient of the fitted straight line in the linear-log plot of $S(t)$ against $t$.

Finally we compare $H$ and $\delta$. For fractional Brownian motion the scaling exponent $\delta$ coincides with the $H$. For random noise with finite variance, the diffusion distribution $p(x,t)$ will converge, according to the central limit theorem, to a Gaussian distribution with $H= \delta =0.5$. If $H \neq \delta$ the scaling represents anomalous behavior. The diffusion processes characterized by Levy flights fall into the class of anomalous diffusion. In this case the Eq. (1) still holds true but the variance is not finite and, therefore, the variance scaling exponent cannot be defined. Another interesting example of the anomalous diffusion is the case of Levy-walk, which is obtained by generalizing the central limit theorem [62]. In this particular kind of diffusion process the second moment is finite but the scaling exponents $H$ and $\delta$ are found to obey the relation

$$\delta = \frac{1}{3-2H} \qquad (6)$$

instead of being equal [63].

### 3.2. The DEA algorithm

In case of time series analysis we have to deal with a sequence of N numbers of $\xi_i$, with $i=1,…,N$ and therefore adopt a discrete picture. From the data sequence we derive the largest possible number of diffusing trajectories with the method of mobile window of integer length $t$. We choose the integer $t$ fitting the condition $1 \leq t \leq N$. This integer plays the role of the diffusion time. Diffusion trajectory:

$$x(t) \equiv \sum_{i=1}^{t} \xi_i, \qquad t = 1, 2, 3, \ldots, N. \qquad (7)$$

The trajectory is then used to build a series of sub-trajectories $\{x_n(t)\}$ according to the following algorithm

$$x_n(t) \equiv \sum_{i=1}^{t} \xi_{i+n-1}, \qquad n = 1, 2, 3, \ldots, N\text{-}t. \qquad (8)$$

where $x_n(t)$ denotes the position of the $n^{th}$ sub-trajectory at time $t$. For each time there are only $N\text{-}t$ available sub-trajectories because the last available sub-trajectory is made by the last $t$ data, i.e., by $\xi_{N-t+1}, \xi_{N-t+2}, \ldots, \xi_{N-1}, \xi_N$. All trajectories start from the origin $x(t=0) = 0$. As time $t$ increases, the sub-trajectories generate a diffusion process. At each time $t$, it is possible to calculate the variance of the position of the $N\text{-}t$ available sub-trajectories according to the well known variance equation:

$$\Sigma^2(t) = \mathrm{var}(x(t)) = \frac{\sum_{n=1}^{N-t}(x_n(t) - \bar{x}(t))^2}{N - t - 1} \qquad (9)$$

where $\bar{x}(t)$ is the average of the positions of the $N-t$ sub-trajectories at time $t$.

At each time t, it is possible to estimate a pdf $p(x,t)$ that will be used to evaluate the entropy of the diffusion process. For that purpose, we partition the *x*-axis into cells of size $\Delta(t)$. After making the partition, we label the cells. We count how many particles are found in each cells at a given time $t$. We denote the number by $N_i(t)$. Then, we use this number to determine the probability that a particle can be found in the *i-th* cell at time *t*, $p_i(t)$, by means of $p_i(t) \equiv \frac{N_i(t)}{(N-t+1)}$. The entropy of the diffusion process at time $t$ is determined by, $S(t) = -\sum_i p_i(t) \ln[p_i(t)]$. For simplicity, the cell size is assumed to be independent of *t* and is determined by a suitable fraction of the square root of the variance of the fluctuation $\xi_i$.

### 3.3. Results of the scaling analysis

Recently DEA and SDA have been applied to various social, meteorological, economical and biological time series data to reveal the exact scaling nature [54-57, 64]. Here we take the initiative to apply these two methods for the time series data of the stress drop of the PLC effect in Al-2.5%Mg alloy to find the exact scaling. Fig. 2 and Fig. 3 show the plot of $D(t)$ and $S(t)$ against $t$ respectively calculated from stress vs. time data taken at $3.85 \times 10^{-4}$ Sec$^{-1}$ strain rate. These plots are fitted with Eqs. (2) and (3) respectively yielding the scaling exponents $H = 0.96$ and $\delta = 0.91$.

The scaling exponents $H$ and $\delta$ obtained from the time series data for different strain rates are listed in Table 1. The high values of the scaling exponents imply a strong persistence in the stress fluctuations. The values of $H$ and $\delta$ decrease marginally due to increase in strain rate. High strain rate demands higher average dislocation velocity and lesser waiting time, causing decrease in the additional activation enthalpy $\Delta G$ due to solute concentration accumulated at the glide dislocation segments. This enhances the probability of the thermally-activated dislocation glide, resulting more load drops in unit time. This causes $H$ and $\delta$ to decrease with increase in strain rate. Finally, we note that the standard deviation scaling exponents ($H$) are larger than the corresponding diffusion entropy scaling exponents ($\delta$) and seen to fulfill the Levy-walk diffusion relation (Eq. (4)) within the accuracy of our statistical analysis as shown in the column 4 of Table 1.

It is evident from different studies [65, 66] that the physics governing the PLC effect necessarily localizes the deformation in the form of bands at all plastic strain level after the critical strain at which PLC effect initiates. Metals deform through the generation and propagation of dislocations, in the sub-micron scale. The key to the PLC effect lies ultimately bundled in the dynamics that connect the microscopic world of dislocations to the macroscopic regime of the bands. Deformation bands travel with constant velocity at constant stress [67]. If the pulling speed is increased, the band velocity changes proportionally. Initial motion of the band is accompanied by a sudden drop in stress. For low strain rate, this drop is sufficient to stop the band. At high strain rates, the stress drops are relatively small in magnitude and no longer stop the band completely and it travels from hopping to constant velocity with the increase in the strain rate. Depending on temperature and strain rate conditions these bands may or may not be correlated in space. This correlation arises due to the long range elastic interactions. In the earlier works [43, 45] it was shown that the dynamics of the PLC bands at low and medium strain rates are chaotic while at high strain rate it showed SOC. But our results from DEA and SDA (Table 1) clearly suggest that the scaling behavior of the overall dynamics of the PLC effect at all strain rates follow Levy-walk property. The strongly correlated glide of macroscopic dislocation groups in the band, the long range elastic interactions among the dislocations and the DSA are the basic ingredients for the dynamics of the PLC bands. The variations of the degree of these three attributes manifest different observable macroscopic dynamic characteristics of the bands. Identicality of the basic dynamical features responsible for the PLC bands, made them to scale uniformly.

## 4. Quantification of Complexity of the PLC effect

A system consisting of many parts which are connected in a nonlinear fashion is designated as the complex system. In a complex system, the interaction between the parts allows the emergence of global behaviour that would not be anticipated from the behaviour of components in isolation. In this respect the PLC effect falls into the class of complex system which posses a real threat to deal it properly whereas understanding its behaviour offers possibility of spectacular and unforeseen advances in many areas of science and its application. It is this threat and this promise that is making complex systems science one of the fastest growing areas of science at the present time. Recently there have been some attempts to quantify the complexity of the complex systems [68-70]. Though there is no formal definition of complexity, it is considered to be a measure of the inherent difficulty to achieve the desired understanding. Simply stated, the complexity of a system is the amount of information necessary to describe it.

To the best of our knowledge no effort has been made to quantify the complexity of the PLC effect observed at different strain rates. Quantifying the complexity of the time series from a physical process may be of considerable interest due to its potential application in evaluating a dynamical model of the system. In this section, we present a quantitative study of the measure of complexity of the PLC effect at the different strain rates. Entropy based algorithm are often used to quantify the regularity of a time series [71, 72]. Increase in entropy corresponds to the increase in the degree of disorder and for a completely random system it is maximum. Traditional algorithms are single-scale based [71, 72]. However, time series derived from the complex systems are likely to present structure on multiple temporal scales. In contrast, time series derived from a simpler system are likely to present structures on just a single scale. For these reasons the traditional single scale based algorithms often yield misleading quantifications of the complexity of a system.

Recently Costa et al. [73] introduced a new method, Multiscale Entropy (MSE) analysis for measuring the complexity of finite length time series. This method measures complexity taking into account the multiple time scales. This computational tool can be quite effectively used to quantify the complexity of a natural time series. The first multiple scale measurement of the complexity was proposed by Zhang [74]. Zhang's method was based on the Shannon entropy which requires a large number of almost noise free data. On the contrary, the MSE method uses Sample Entropy (SampEn) to quantify the regularity of finite length time series. SampEn is largely independent of the time series length when the total number of data points is larger than approximately 750 [75]. Thus MSE proved to be quite useful in analyzing the finite length time series over the Zhang's method. Recently MSE has been successfully applied to quantify the complexity of many Physiologic and Biological signals [73, 76, 77]. Here, we take the initiative to apply this novel method to quantify the complexity of a metallurgical phenomenon.

To study the complexity of the PLC effect in three different strain rate regime we have taken the data for the strain $7.56 \times 10^{-6}$ sec$^{-1}$, $1.99 \times 10^{-4}$ sec$^{-1}$ and $1.92 \times 10^{-3}$ sec$^{-1}$. These strain rates were chosen in such a way that we could observe only type C, type B and type A serrations respectively in the stress time data [47]. The stress time data obtained from these experiments show an increasing drift due to the strain hardening effect. To remove the effect of strain hardening we have subtracted the drift by the method of polynomial fitting. The investigations presented below were carried out on the drift corrected data. The segments of the drift corrected stress time curves for the three strain rates are shown in Fig. 4.

### 4.1. Multiscale Entropy Analysis

The MSE method is based on the evaluation of SampEn on the multiple scales. The prescription of the MSE analysis is: given a one-dimensional discrete time series, $\{\xi_1,\ldots,\xi_i,\ldots,\xi_N\}$, construct the consecutive coarse-grained time series, $\{y^{(\tau)}\}$, determined by the scale factor, $\tau$, according to the equation:

$$y_j^\tau = 1/\tau \sum_{i=(j-1)\tau+1}^{j\tau} \xi_i \qquad (10)$$

where $\tau$ represents the scale factor and $1 \leq j \leq N/\tau$. The length of each coarse-grained time series is $N/\tau$. For scale one, the coarse-grained time series is simply the original time series. Next we calculate the SampEn for each scale using the following method. Let $\{X_i\} = \{\xi_1,\ldots,\xi_i,\ldots,\xi_N\}$ be a time series of length N. $u_m(i) = \{\xi_i, \xi_{i+1},\ldots,\xi_{i+m-1}\}, 1 \leq i \leq N-m$ be vectors of length $m$. Let $n_{im}(r)$ represent the number of vectors $u_m(j)$ within distance $r$ of $u_m(i)$, where $j$ ranges from 1 to (N-m) and $j \neq i$ to exclude the self matches. $C_i^m(r) = n_{im}(r)/(N-m-1)$ is the probability that any $u_m(j)$ is within $r$ of $u_m(i)$. We then define

$$U^m(r) = 1/(N-m) \sum_{i=1}^{N-m} \ln C_i^m(r) \qquad (11)$$

The parameter Sample Entropy (SampEn) [75] is defined as

$$SampEn(m,r) = \lim_{N \to \infty}\left\{-\ln\frac{U^{m+1}(r)}{U^m(r)}\right\} \qquad (12)$$

For finite length N the SampEn is estimated by the statistics

$$SampEn(m,r,N) = -\ln\frac{U^{m+1}(r)}{U^m(r)} \qquad (13)$$

Advantage of SampEn is that it is less dependent on time series length and is relatively consistent over broad range of possible r, m and N values. We have calculated SampEn for all the studied data sets with the parameters m=2 and r= 0.15×SD (SD is the standard deviation of the original time series).

### 4.2. Results of the MSE analysis

Costa et al. had tested the MSE method on simulated white and 1/f noises [73]. They have shown that for the scale one, the value of entropy is higher for the white noise time series in comparison to the 1/f noise, as shown in Fig. 5. This may apparently lead to the conclusion that the inherent complexity is more in the white noise in comparison to the 1/f noise. However, the application of the MSE method shows that the value of the entropy for the 1/f noise remains almost invariant for all the scales while the value of entropy for the white noise time series monotonically decreases and for scales > 5, it becomes smaller than the corresponding values for the 1/f noise. This result explains the fact that the 1/f noise contains complex structures across multiple scales in contrast to the white noise. With a view to understand the complexity of a chaotic process we have generated chaotic data from the logistic map $\xi_{n+1}= \xi_n (1-\xi_n)$ with a=3.9 and applied the MSE method. The entropy measure for the chaotic time series increases on small scales and then gradually decreases indicating the reduction of complexity on the larger scales, as shown in Fig. 5.

We next apply the MSE method to the analysis of the stress fluctuations observed in the PLC effect. The results of the MSE analysis of the stress time series for the three studied strain rates are shown in Fig. 6. It is seen that for the lowest strain rate, where only the type C bands are observed, the SampEn is very low for all the scales. The entropy measure slightly increases with the scale. This signifies the simplicity of the type C band dynamics compared to the other two bands. For the medium strain rate i.e. for the type B serrations, the entropy measure markedly increases in the small scales and then gradually decreases and gets saturated. The entropy measure for the high strain rate stress time series (type A serrations) decreases on the small scale and then gradually increases. For scale one, the entropy assigned to the stress time series for the type A serrations is higher than that of the type B and C stress time series. In contrast for scales > 3 the time series of type B serrations are assigned to the highest entropy values. For the largest scale, the entropy measures for the type B and the type A serrations become almost equal. Moreover, for the type B serrations the variation of the SampEn with the scale is similar to that of the chaotic data [Fig. 2]. This result corroborates with the previous finding of the presence of deterministic chaos in the type B band dynamics [78, 79]. The high values of entropy for all the scales for the type A stress time series data may be a signature of the SOC as observed earlier [44].

A possible approach for understanding the observed complexity measure is to consider the properties of the time and length scales relevant to the PLC effect. At low strain rates, the reloading time between two successive drops, $t_l$, is very large compared to the plastic relaxation time $t_r$ i.e. $t_l \gg t_r$. The internal stresses are fully relaxed in this case and consequently the type C bands are spatially uncorrelated. The absence of spatial correlation makes the dynamics of the type C band simple and it may be easier to build a model based on the repeated pinning and unpinning of dislocations in the field of solute atoms. When the driving strain rate is increased, the reloading time scale $t_l$ decreases and approaches the relaxation time scale $t_r$. Internal stresses are not totally relaxed during the reloading periods. Hence, new bands are formed nearby the previous ones leading to the hopping character of the associated type B bands. This spatial coupling and the competition between the two time scales increases the complexity of the type B band dynamics. At high strain rates the ratio $t_l/t_r$ decreases, which hinders the plastic relaxation. Very little plastic relaxation occurs between the stress drops. Thus, the stress felt by the dislocations always remain close to the threshold for unpinning from the solute atmosphere. New bands are formed in the field of the unrelaxed internal stresses and perennial plastic events overlap resulting in a hierarchy of length scales. This leads to both SOC dynamics and type A propagating bands [43]. Due to this hierarchy the entropy measure for the high strain rate comes out to be high for all the scales. The high complexity of the type B and A band dynamics makes the modeling of PLC effect very difficult.

In recent years there have been a great enthusiasm to model the PLC effect. Some good models [30, 33, 80], which reproduce some of the features of the PLC effect very well have also been emerged. However, there are constant efforts to provide better models. Knowledge of this complexity measure should be taken care of while modeling the PLC effect in different strain rate region.

## 5. Conclusions

In conclusion, we have studied the PLC effect from a new perspective and evaluated the exact scaling behavior of the PLC effect in Al-2.5%Mg alloy using two complementary scaling analysis methods: DEA and SDA. The analyses were performed in a wide range of strain rates where different types of deformation bands are observed. The relation of the two scaling exponents in each strain rate obtained through our analysis clearly suggests that the stress fluctuations occurring due to dislocation flow (PLC effect) in Al-2.5%Mg alloy inherit a Levy-walk memory component and the scaling behavior is independent of strain rate.

We have also quantified the complexity of the PLC effect at different strain rate regime. We have carried out the Multiscale Entropy analysis on the PLC effect. The entropy measure is very low for the low strain rate i.e. for type C serrations, signifying the simplicity of the type C band dynamics. For medium strain rate the entropy increases markedly in the small scales and then gradually decreases and gets saturated. For the high strain rate the entropy decreases in small scales and then gradually increases with the scale and finally at the largest scale it becomes almost equal to the entropy measure of the medium strain rate. The study clearly establishes the fact that the PLC dynamics of the type A and B bands is the most complex one.

## Acknowledgement

The authors thank Dr. N. Scafetta and Dr. M. Costa for helpful discussions.

# References

[1] G. I. Taylor, Proc. roy. Soc. A 145 (1934) 362.

[2] E. Z. Orowan, Z. Phys. 89 (1934) 634.

[3] M. Polanyi, Ibid 89 (1934) 660.

[4] J. M. Hedges, J. W. Mitchell, Philos. Mag. 44 (1953) 223.

[5] A.H. Cottrell, Dislocations and plastic flow in crystals (Oxford University Press, London, 1953).

[6] J.D. Baird, The Inhomogeneity of Plastic Deformation (American Society of Metals, OH, 1973).

[7] A. Van den Beukel, Physica Status Solidi(a) 30 (1975) 197.

[8] A.W. Sleeswyk, Acta Metall. 6 (1958) 598.

[9] P.G. McCormick, Acta Metall. 20 (1972) 351.

[10] A. Van den Beukel, U. F. Kocks, Acta Metall. 30 (1982) 1027.

[11] P. Penning, Acta Metall. 20 (1972) 1169.

[12] L.P. Kubin, Y. Estrin, Acta Metall. 33 (1985) 397.

[13] F. Savart, Ann. Chim. Phys. 65 (1837) 337.

[14] A. M. Masson, Ann. Chim. Phys. 3 (1841) 451.

[15] F. Le. Chatelier, Rev. de Metall. 6 (1909) 914.

[16] A. Portevin, F. Le. Chatelier, C. R. Acad. Scie., Paris, 176 (1923) 507.

[17] A. H. Cottrell, Phil. Mag. 74 (1953) 829.

[18] A. Wijler, J. Schade van Westrum, Scripta Metall. 5 (1971) 531.

[19] J. M. Robinson, M. P. Shaw, Mat. Sci. Eng. A 159 (1992) 159.

[20] Y. Brechet, Y. Estrin, Acta Metall. Mater. 43 (1995) 955.

[21] P. G. McCormick, Acta Metall. 21 (1973) 873.

[22] R. Shabadi, S. Kumar, H. J. Roven, E. S. Dwarakadasa, Mat. Sci. Eng. A 364 (2004) 140.

[23] H. Louche, P. Vacher, R. Arrieux, Mat. Sci. Eng. A 404 (2005) 188.

[24] Q. Zhang, Z. Jiang, H. Jiang, Z. Chen, X. Wu, International Journal of Plasicity 21 (2005) 2150.

[25] W. Tong, H. Tao, N. Zhang, L. G. Hector Jr., Scripta Mater. 53 (2005) 87.

[26] B.J. Brindley, P.J. Worthington, Metall. Rev. 145 (1970) 101.

[27] L.J. Cuddy, W.C. Leslie, Acta Metall. 20 (1972) 1157.

[28] K. Chihab, Y. Estrin, L.P. Kubin, J. Vergnol, Scripta Metall. 21 (1987) 203.

[29] A. Kalk, Ch. Schwink, Phil. Mag. A 72 (1995) 315.

[30] M. Zaiser, P. Hahner, Physica Status Solidi(b) 199 (1997) 267.

[31] E. Pink, A. Grinberg, Acta MEtall. 30 (1982) 469.

[32] P. Rodriguez, Bull. Mater. Sci. 6 (1984) 653.

[33] L.P. Kubin, C. Fressengeas, G. Ananthakrishna, Dislocations in Solids, Vol. 11, ed. F. R. N. Nabarro, M. S. Duesbery (Elsevier Science, Amsterdam, 2002, p 101).

[34] M. S. Bharathi, Spatio-temporal Dynamics of the Portevin-Le Chatelier Effect: PhD Thesis, 2002.

[35] P. Hahner, Mat. Sci. Eng A 164 (1993) 23.

[36] P. Hahner, A. Ziegenbein, E. Rizzi, H. Neuhauser, Phys. Rev. B 65 (2002) 134109.

[37] G. Ananthakrishna, M.C. Valsakumar, J. Phys. D 15 (1982) L171.

[38] V. Jeanclaude, C. Fressengeas, L.P. Kubin, Nonlinear Phenomena in Materials Science II, ed. L.P. Kubin (Trans. Tech., Aldermanndorf, 1992, p. 385).

[39] M. Lebyodkin, Y. Brechet, Y. Estrin, L.P. Kubin, Acta Mater. 44 (1996) 4531.

[40] M. Lebyodkin, L. Dunin-Barkowskii, Y. Brechet, Y. Estrin, L.P. Kubin, Acta Mater. 48 (2000) 2529.

[41] M.A. Lebyodkin, Y. Brechet, Y. Estrin, L.P. Kubin, Phys. Rev. Lett. 74 (1995) 4758.

[42] M.S. Bharathi and G. Ananthakrishna, Phys. Rev. E. 67 (2003) 065104.

[43] M.S. Bharathi, M. Lebyodkin, G. Ananthakrishna, C. Fressengeas, L.P. Kubin, Acta Mater. 50 (2002) 2813.

[44] G. Ananthakrishna, S. J. Noronha, C. Fressengeas and L. P. Kubin, Phys. Rev. E 60, (1999) 5455.

[45] M.S. Bharathi, M. Lebyodkin, G. Ananthakrishna, C. Fressengeas, L.P. Kubin, Phys. Rev. Lett. 87 (2001) 165508.

[46] S. Kok, M.S. Bharathi, A.J. Beaudoin, C. Fressengeas, G. Ananthakrishna, L.P. Kubin, M. Lebyodkin, Acta Mater. 51 (2003) 3651.

[47] K. Chihab, C. Fressengeas, Mater. Sci. Eng. A 356 (2003) 102.

[48] H.E. Stanley, L.A.N. Amaral, P. Gopikrishnan, P.Ch. Ivanov, T.H. Keitt, V. Plerou, Physica A 281 (2000) 60.

[49] B.B. Mandelbort, The Fractal Geometry of Nature, W.H.Freeman, New York, 1982.

[50] P. Bak, K. Christensen, L. Danon, T. Scanlon, Phys. Rev. Lett. 88 (2002) 178501.

[51] A.P. Siebesma, H.J.J. Jonker, Phys. Rev. Lett. 85 (2000) 214.

[52] S.N. Dorogovtsev, J.F.F. Mendes, Phys. Rev. E 63 (2001) 056125.

[53] G. Zumofen, J. Klafter, M. F. Shlesinger, Phys. Rep. 290 (1997) 157.

[54] N. Scafetta, P. Hamilton, P. Grigolini, Fractals 9 (2001) 193.

[55] N. Scafetta, P. Grigolini, Phys. Rev. E 66 (2002) 036130.

[56] N. Scafetta, V. Latora, P. Grigolini, Phys. Lett. A 299 (2002) 565.

[57] P. Allegrini, V. Benci, P. Grigolini, P. Hamilton, M. Ignaccolo, G. Menconi et. al. , Chaos, Solitions & Fractals 15 (2003) 517.

[58] C.K. Peng, S.V. Buldyrev, S. Havlin, M. Simons, H.E. Stanley, A.L. Goldberger, Phys. Rev. E 49 (1994) 1685.

[59] N. Scafetta, B.J. West, Phys. Rev. Lett. 92 (2004) 138501.

[60] P. Grigolini, D. Leddon, N. Scafetta, Phys. Rev. E 65 (2002) 046203.

[61] N. Scafetta, V. Latora, P. Grigolini, Phys. Rev. E 66 (2002) 031906.

[62] B.V. Gnedenko, A.N. Kolomogorov, Limit Distributions for Sums of Random Variables (Addision-Wesley, MA, 1954).

[63] N. Scafetta, An Entropic Approach To The Analysis of Time Series: PhD Dissertation, 2001

[64] A. Sarkar, P. Barat, Physica A 364 (2006) 362.

[65] E. Rizzi, P. Hahner, Int. J. Plasticity 20 (2004) 121.

[66] S.V. Franklin, F. Mertens, M. Marder, Phys. Rev. E 62 (2000) 8195.

[67] F. Mertens, S.V. Franklin, M. Marder, Phys. Rev. Lett. 78 (1997) 4502.

[68] P. Grassberger, Physica 140A (1986) 319.

[69] J. S. Shiner, M. Davison, Phys. Rev. E 59 (1999) 1459.

[70] C. Bandt, B. Pompe, Phys. Rev. Lett. 88 (2002) 174102.

[71] S. M. Pincus and A. L. Goldberger, Am. J. Physiol. Heart Circ. Physiol. 266 (1994) H1643.

[72] R. Gunther, B. Shapiro and P. Wagner, Chaos, Solitons and Fractals 4 (1994) 635.

[73] M. Costa, A. L. Goldberger and C. −K. Peng, Phys. Rev. Lett. 89 (2002) 068102.

[74] Y. –C. Zhang, J. Phys. I(France) 1 (1991) 971.

[75] J. S. Richman and J. R. Moorman, Am. J. Physiol. 278 (2000) H2039.

[76] M. Costa, A. L. Goldberger and C. –K. Peng, Phys. Rev. E. 71 (2005) 021906.

[77] M. Costa, C. –K. Peng, A. L. Goldberger and J. M. Hausdorff, Physica A 330 (2003) 53.

[78] S. J. Noronha, G. Ananthakrishna, L. Quaouire, C. Fressengeas and L. P. Kubin, Int. J. Bifurcation Chaos Appl. Sci. Eng. 7 (1997) 2577.

[79] G. Ananthakrishna, C. Fressengeas and L. P. Kubin et.al. Scripta Metall. Mater. 32 (1995) 1731.

[80] Y. Estrin, L.P. Kubin, Continuum Models for Materials with Microstructure, ed. H.B. Muhlhaus (Wiley, New York, 1995, p. 395).

TABLE 1. SDA and DEA scaling exponents obtained from the Stress vs. Time data of Al-2.5%Mg alloy during tensile deformation for different strain rates.

| Strain rate ($Sec^{-1}$) | SDA exponent ($H$) (Maximum error=±0.02) | DEA exponent ($\delta$) (Maximum error=±0.02) | $\left[\left(\delta - \dfrac{1}{3-2H}\right)/\delta\right] \times 100$ |
|---|---|---|---|
| $7.56 \times 10^{-6}$ | 0.98 | 0.94 | 2.29 |
| $1.99 \times 10^{-5}$ | 0.98 | 0.93 | 3.39 |
| $3.98 \times 10^{-5}$ | 0.97 | 0.92 | 2.54 |
| $7.99 \times 10^{-5}$ | 0.96 | 0.92 | 0.64 |
| $1.55 \times 10^{-4}$ | 0.96 | 0.92 | 0.64 |
| $3.85 \times 10^{-4}$ | 0.96 | 0.91 | 1.75 |
| $5.88 \times 10^{-4}$ | 0.96 | 0.91 | 1.75 |
| $7.42 \times 10^{-4}$ | 0.96 | 0.91 | 1.75 |
| $1.21 \times 10^{-3}$ | 0.95 | 0.90 | 1.01 |
| $1.59 \times 10^{-3}$ | 0.94 | 0.89 | 0.32 |
| $1.92 \times 10^{-3}$ | 0.94 | 0.88 | 1.46 |

**Figure captions**

Fig. 1. True Stress vs. True Strain curve of Al-2.5%Mg alloy at a strain rate of $3.85\times10^{-4}$ Sec$^{-1}$. The inset shows a typical region of the curve in the Stress-Time plot.

Fig. 2. SDA of the Stress vs. Time data obtained from Al-2.5%Mg alloy during tensile deformation at a strain rate of $3.85\times10^{-4}$ Sec$^{-1}$.

Fig. 3. DEA of the Stress vs. Time data obtained from Al-2.5%Mg alloy during tensile deformation at a strain rate of $3.85\times10^{-4}$ Sec$^{-1}$.

Fig. 4. Segments of the drift corrected stress time curves at three strain rates: (a) $7.56\times10^{-6}$ sec$^{-1}$, (b) $1.99\times10^{-4}$ sec$^{-1}$ and (c) $1.92\times10^{-3}$ sec$^{-1}$.

Fig. 5. MSE analysis of white noise, 1/f noise and logistic map chaotic data. On the y axis, the value of the entropy (SampEn) for the coarse-grained time series is plotted. The scale factor specifies the number of data points averaged to obtain each element of the coarse-grained time series.

Fig. 6. MSE analysis of the stress time series for the strain rates: $7.56\times10^{-6}$ sec$^{-1}$ (low), $1.99\times10^{-4}$ sec$^{-1}$ (medium), and $1.92\times10^{-3}$ sec$^{-1}$ (high).

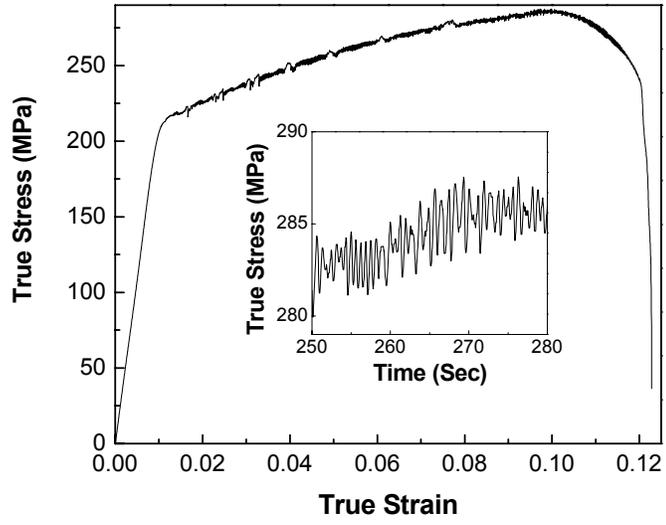

Fig. 1

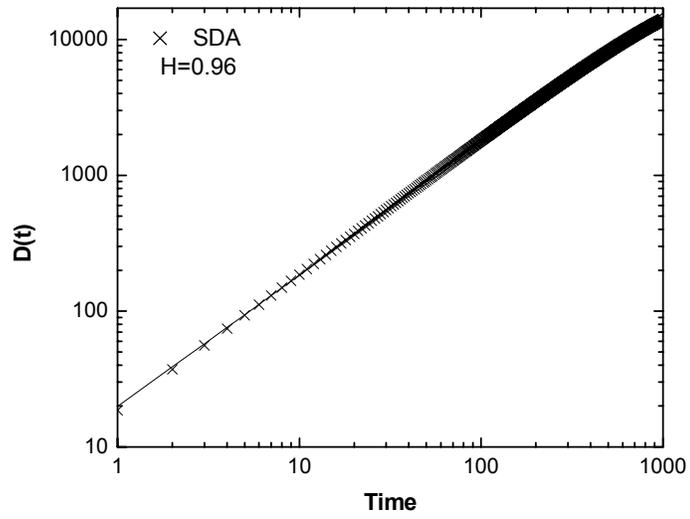

Fig. 2

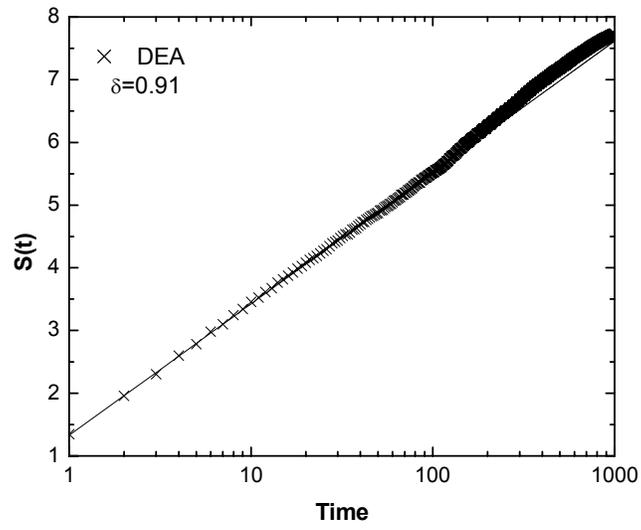

Fig. 3

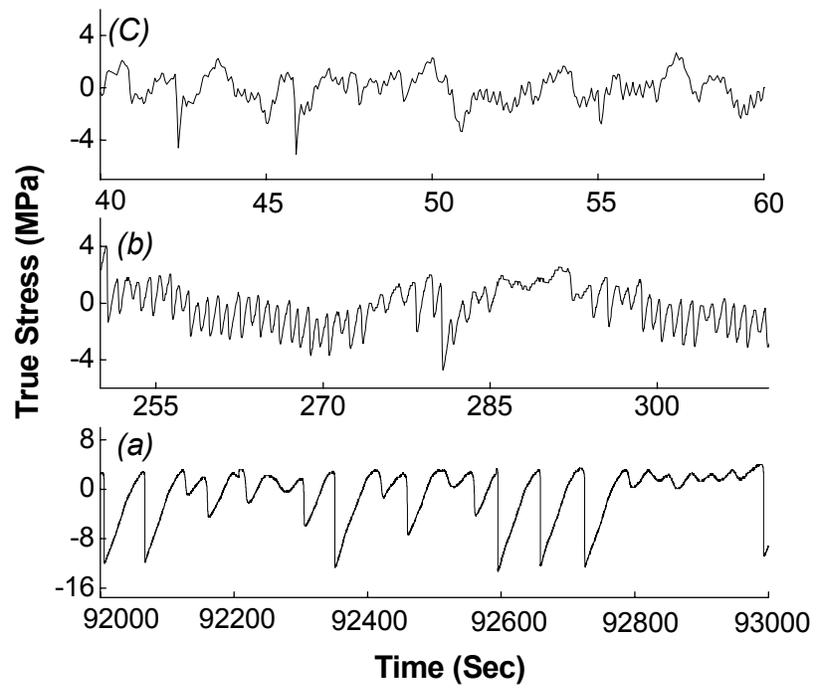

Fig. 4

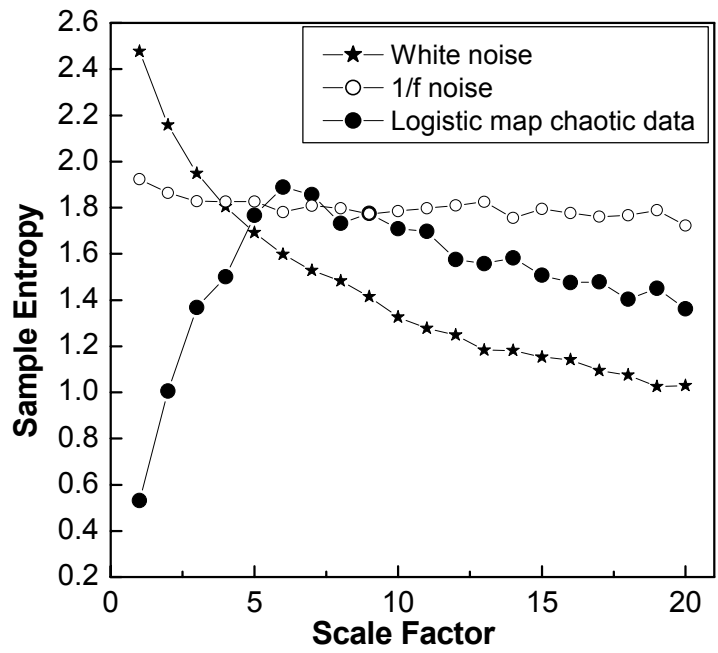

Fig. 5

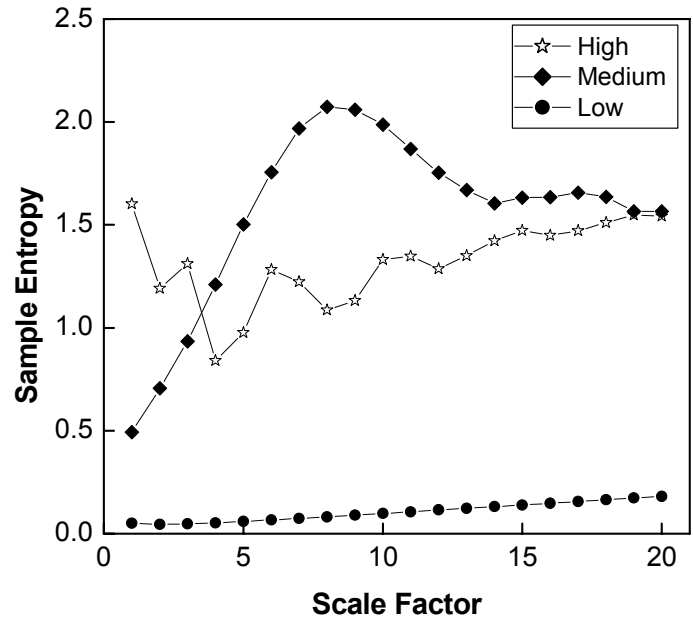

Fig. 6